# BLOCKCHAIN AND STABLECOIN INTEGRATION FOR CROWDFUNDING: A FRAMEWORK FOR ENHANCED EFFICIENCY, SECURITY, AND LIQUIDITY

**Mustafa Savas, Ünsal**
*Department of Management Information Systems*
*Faculty of Economics, Administrative And Social Sciences*
*Istanbul Nisantasi University,*
*Tasyoncası Street, 34398, Sarıyer, Istanbul, Türkiye*
mustafasavas.unsal@nisantasi.edu.tr

January 19, 2025

## ABSTRACT

Crowdfunding platforms face high transaction fees, need for more transparency, and trust deficits. These issues deter contributors and entrepreneurs from effectively leveraging crowdfunding for innovation and growth. Blockchain technology introduces decentralization, security, and efficiency to address these limitations (1). This paper proposes a blockchain-based crowdfunding framework that integrates stablecoins such as USDT and USDC to mitigate cryptocurrency volatility and ensure seamless fund management. Smart contracts automate compliance processes, including Know Your Customer (KYC) / Anti-Money Laundering (AML) checks, and enhance operational efficiency (2). Furthermore, tokenization enables liquidity by allowing fractional ownership and secondary market trading, which must be effectively implemented on any global market platform. A comparative analysis highlights the superiority of the framework over traditional platforms in terms of cost reduction, transparency, and investor trust. A case study focused on the Turkish market illustrates the practical benefits of blockchain adoption in equity crowdfunding, particularly in navigating local regulatory and financial complexities. This approach provides a scalable, secure, and accessible solution for modern crowdfunding ecosystems, while reducing the costs of platforms and increasing the trust of investors and backers in crowdfunding projects.

## 1. Introduction

Crowdfunding has revolutionized the entrepreneurial landscape, bridging the gap between creators and a diverse pool of investors (3). Platforms like Kickstarter, Indiegogo, and Crowdcube have enabled billions in capital to flow directly to entrepreneurs by bypassing traditional financial institutions (4). Despite its success, the model faces persistent challenges, including high transaction fees, lack of transparency, regulatory hurdles, and scalability issues.

Traditional crowdfunding platforms depend on intermediaries like payment gateways and escrow services, which charge 3% to 5% fees, significantly reducing the funds creators receive. Transparency is another critical issue, as contributors often lack insights into fund allocation, discouraging repeat investments. Regulatory compliance, particularly AML and KYC requirements, introduces delays and increases operational costs. Furthermore, centralized infrastructures limit scalability, hindering large campaigns and global audiences (5).

Blockchain technology addresses these inefficiencies through its decentralized, transparent, and secure framework. It reduces transaction costs by eliminating intermediaries and uses immutable ledgers to enhance trust and accountability. Blockchain also mitigates fraud risks, allowing investors to monitor projects in real-time.

While cryptocurrencies like Bitcoin face adoption barriers due to volatility, stablecoins such as USDT and USDC provide a practical alternative (6). These assets maintain stable values, facilitating predictable transactions and seamless conversion into local currencies, particularly in emerging markets like Turkey.

This paper introduces a blockchain-based crowdfunding framework that integrates tokenization, smart contracts, and stablecoins to improve transparency, reduce costs, and improve liquidity.

## 2. Proposed Framework

The framework leverages Blockchain's decentralized ledger technology to address inefficiencies in traditional crowdfunding platforms, building on its proven applications beyond cryptocurrency (7). Smart contracts utilize Blockchain's immutable and decentralized nature to automate compliance and fund management processes, as Pilkington (8) described. These features ensure secure and efficient fund distribution, milestone tracking, and refunds for contributors.

The proposed blockchain-based crowdfunding framework integrates stablecoin transactions, tokenization, and smart contracts into a decentralized architecture. By addressing the inefficiencies of traditional crowdfunding platforms, this framework aims to reduce costs, enhance transparency, and provide liquidity to contributors.

### 2.1. Stablecoin Integration

Stablecoins like USDT and USDC mitigate the volatility of cryptocurrencies such as Bitcoin or Ether (6). Their stability, achieved through a 1:1 peg to fiat currencies, provides a reliable financial medium for crowdfunding platforms. Using stablecoins enables cost-efficient transactions by eliminating intermediaries like banks and credit card processors. Platforms like BtcTurk further enhance usability by allowing seamless conversion into local currencies such as the Turkish Lira (TRY).

The workflow for stablecoin integration begins with contributors sending stablecoins directly to a project's wallet. The platform monitors these transactions through Blockchain-based ledgers, ensuring full transparency and immutability. Stablecoins received by project owners are then converted to fiat currency as required, minimizing financial barriers for local stakeholders. The framework ensures low-cost, transparent, and stable financial operations by replacing traditional payment systems with stablecoins.

### 2.2. Tokenization

Tokenization is the digital representation of project rights or assets on a Blockchain (9). While tokenization has gained prominence for its flexibility and liquidity, overemphasizing these aspects often obscures other potential advantages. Through this framework, tokenization enhances transparency by recording all transactions on an immutable ledger. Contributors can monitor their ownership rights in real time, fostering trust and accountability.

Tokenization also enables fractional ownership, democratizing investment access by allowing contributors to own small portions of a project. For example, in a real estate crowdfunding campaign, tokens can represent a fraction of a property's ownership. These tokens are tradable on secondary markets, offering investors liquidity that traditional crowdfunding platforms cannot provide. Additionally, the framework supports various token designs, such as equity tokens for ownership, reward tokens for specific perks, and hybrid tokens combining multiple features. By leveraging tokenization, the framework attracts a wider pool of contributors while ensuring that funds remain liquid and traceable.

Smart contracts form the backbone of the proposed framework, automating compliance and fund management processes (2). These self-executing agreements reduce administrative overhead and enhance security by eliminating manual intervention. For instance, funds are locked into escrow smart contracts and are only released when predefined milestones are met. This milestone-based payment system ensures that project owners remain accountable to contributors.

Smart contracts also simplify compliance processes. By embedding KYC / AML protocols directly into the contract's logic, the framework ensures adherence to regulatory standards without introducing delays. Refund mechanisms are another critical feature of smart contracts. They automatically return funds to contributors if a campaign fails to meet its funding goal. This automation builds trust among stakeholders by removing ambiguities in fund allocation. A solidity-based implementation highlights the technical underpinnings of these contracts. As an example;

```
function refund() public {
    require(block.timestamp >= deadline, "Funding period still active");
    uint256 amount = contributions[msg.sender];
    contributions[msg.sender] = 0;
    payable(msg.sender).transfer(amount);
}
```

Here the **key components** listed as follows:

> **function refund() public{**
> *It declares a public function named refund, which anyone can call.*
>> **require(block.timestamp >= deadline, "Funding period still active");**
>> *Ensures the current timestamp (block.timestamp) is greater than or equal to a predefined deadline.*
>> *If the condition is unmet, the transaction reverts to the message "Funding period still active".*
>> **uint256 amount = contributions[msg.sender];**
>> *Fetches the amount contributed by the caller (msg.sender) from the contributions mapping.*
>> **contributions[msg.sender] = 0;**
>> *Resets the caller's contribution to 0 to prevent double refunds.*
>> **payable(msg.sender).transfer(amount);**
>> *Transfers the contributed amount back to the caller (msg.sender).*
>> *Payable ensures that the address can receive USDT / USDC.*
>
> **}**

This code demonstrates how contributors are refunded automatically if project deadlines are unmet.

### 2.3. Architecture Overview

The architecture integrates stablecoins, smart contracts, and tokenization within a decentralized system. Contributors use secure wallets to fund projects and receive tokens, while project creators manage campaigns through an intuitive interface. Transactions are recorded on a Blockchain network to ensure transparency. Regulatory nodes monitor compliance with local laws, while milestones dictate fund disbursement. This architecture ensures scalability, security, and seamless interaction between stakeholders. The architecture diagram below illustrates the interactions between contributors, project creators, smart contracts, and regulatory nodes:

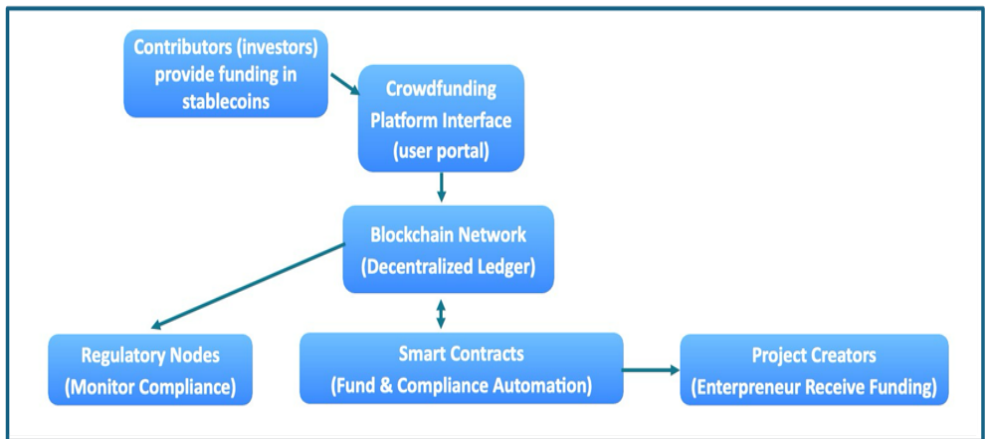

Figure 1: Blockchain-Based Crowdfunding Architecture Diagram (created by author).

Legend:

Arrows indicate fund flow and data exchanges between components.

Circular nodes represent decentralized points of interaction.

## 3. Methodology

The proposed blockchain-based crowdfunding framework integrates stablecoins, tokenization, and smart contracts to address inefficiencies in traditional platforms. This section outlines the structured workflow, compliance mechanisms, and security measures that ensure scalability and accountability.

### 3.1. Workflow

The framework operates through a detailed workflow designed to automate processes and ensure transparency. Entrepreneurs initiate campaigns by defining the project scope, funding goals, milestones, and rewards. Predefined rules encoded in smart contracts validate the campaign, issuing tokens representing equity, rewards, or project ownership. Contributors access the platform via secure wallets to fund campaigns using stablecoins such as USDT or USDC (6). Smart contracts immediately acknowledge contributions, issuing corresponding tokens to contributors. Funds are stored in escrow-like smart contracts and disbursed incrementally based on milestone completion, verified by third-party auditors or validators.

Tokens issued during the campaign are tradable on blockchain-based secondary markets, providing contributors with liquidity. Upon project completion, contributors receive rewards or profit shares as encoded in smart contracts, and the remaining funds are transferred to the campaign owner.

### 3.2. Tokenization

Tokenization is integral to the platform, enabling liquidity, transparency, and accessibility. Tokens are created and distributed based on predefined campaign structures and funding contributions. For instance, a contributor providing 1% of the total funding receives 1% of the allocated tokens.

These tokens can be traded on secondary markets or utilized in Decentralized Finance (DeFi) platforms for staking or collateral, enhancing their value and usability. The framework ensures fairness and traceability by integrating tokenization with smart contracts, simplifying fund management while democratizing investment opportunities.

### 3.3. Compliance

The framework enforces regulatory compliance through automated processes embedded in smart contracts (2). KYC / AML rules verify the identity of contributors and campaign creators, barring non-compliant users from participation. Regulatory nodes continuously monitor transactions, generating real-time reports to streamline auditing and reduce manual reporting costs.

The framework ensures cross-border compatibility by embedding localized regulatory requirements, enabling seamless operation across jurisdictions.

### 3.4. Security

Security is a cornerstone of the methodology. Immutable smart contracts deployed on the Blockchain ensure integrity and fairness. Contributor and campaign data are encrypted and stored off-chain, linking only necessary data points to the Blockchain to maintain privacy. Fraud prevention is achieved through milestone-based payouts and real-time monitoring of transactions.

Multi-signature wallets require approvals from multiple stakeholders for critical actions, reducing the risk of unauthorized transactions. These measures ensure the safety and reliability of the platform.

### 3.5. Scalability

The platform achieves scalability through horizontal scaling and efficient consensus mechanisms like Proof-of-Stake (PoS) or Delegated Proof-of-Stake (DPoS), enabling rapid transaction validation with minimal energy consumption. Stablecoin integration eliminates the need for complex exchange mechanisms, further streamlining operations and ensuring high performance during periods of increased activity.

The following flowchart illustrates the complete methodological process:

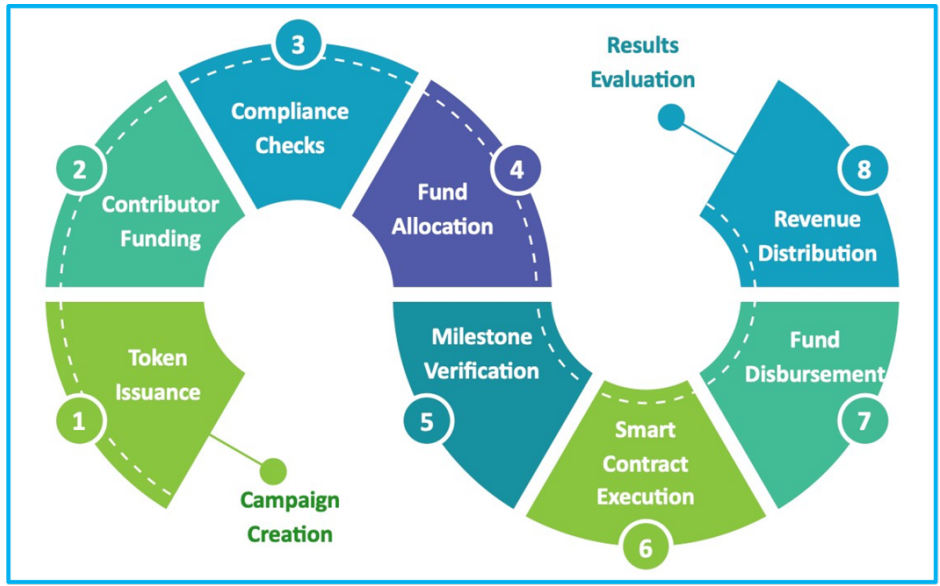

Figure 2: Blockchain-Based Crowdfunding Framework (created by author).

## 4. Comparative Analysis

Crowdfunding platforms have significantly evolved, transitioning from traditional centralized models to innovative Blockchain-based solutions. This section compares cost, transparency, security, and scalability approaches.

### 4.1. Cost Efficiency

Traditional platforms incur high transaction fees, typically 5% to 10% of funds raised. These fees encompass platform commissions, payment gateway charges, and intermediary costs. Blockchain-based platforms, on the other hand, leverage smart contracts and stablecoins to reduce fees significantly, often to less than 3% of the total funds raised (10). By eliminating intermediaries, the Blockchain framework ensures that a larger share of contributions directly supports campaign objectives.

### 4.2. Transparency

Centralized crowdfunding platforms often lack transparency, leaving contributors uncertain about fund allocation and project progress. Blockchain-based platforms mitigate this issue through immutable ledgers that record every transaction. Contributors can monitor fund flows and project milestones in real time, fostering stakeholder trust (11). Integrating smart contracts further enhances transparency by automating fund disbursement based on predefined conditions.

### 4.3. Security

Traditional platforms are vulnerable to fraud and data breaches, and security is a significant concern. Blockchain's decentralized architecture and cryptographic principles significantly reduce these risks. Funds are stored in secure, tamper-proof smart contracts, ensuring they are only released upon achieving project milestones. Moreover, multi-signature wallets provide additional protection against unauthorized access (2).

### 4.4. Scalability

Centralized platforms face scalability limitations due to their reliance on intermediaries and centralized servers. Blockchain-based systems overcome these barriers by employing distributed ledger technology, enabling seamless operation even during periods of high transaction volume. Stablecoin integration ensures smooth cross-border transactions, eliminating the need for currency conversions and associated delays.

Table 1: Comparative Analysis

| Feature | Traditional Platforms | Blockchain-Based Framework |
|---|---|---|
| Transaction Fees | High (3-5 %) | Low (<1 %) |
| Transparency | Limited | Complete (Blockchain ledger) |
| Fraud Prevention | Limited | High (Smart contract enforcement) |
| Scalability | Centralized, prone to bottlenecks | Decentralized, highly scalable |
| Liquidty | Locked until project completion | Tradeable tokens on secondary markets |
| Compliance Costs | High (manual checks) | Low (automated via smart contracts) |
| Cross-Border Compatibility | Restricted | Global (Stablecoin integration) |

Blockchain-based solutions provide a cost-efficient, transparent, and secure alternative to traditional crowdfunding platforms by addressing their critical shortcomings. These advancements are particularly beneficial for emerging markets, where inefficiencies in traditional systems have historically hindered participation and growth.

## 5. Case Study: Application in Turkey

The Turkish equity crowdfunding landscape has experienced significant growth following key regulatory milestones in 2017 and 2019. Amendments to Capital Markets Law No. 6362 legitimized equity-based crowdfunding platforms, providing a foundation for structured and regulated fundraising activities. Platforms like Fonbulucu and Fongogo have been pivotal, collectively managing over 56% of the market share in recent years (1).

Despite advancements, the ecosystem faces high operational costs, limited investor trust, and currency volatility. Recent research on crowdfunding project supporters in Turkiye (12) has shown that the reliability and accessibility factor of the crowdfunding platform is an important factor that influences the motivation of supporters to contribute to projects. Supporters may desire the platform to be secure and accessible while contributing to projects. Therefore, crowdfunding platforms must provide a reliable and user-friendly experience. Blockchain technology offers solutions by enhancing transparency, reducing costs, and introducing tokenization and stablecoins as transformative tools (5; 11). Stablecoins like USDT and USDC mitigate currency volatility, making crowdfunding accessible to international investors and establishing Turkey as a potential regional leader in blockchain-enabled equity crowdfunding.

### 5.1. Regulatory Milestones and Equity-Based Platforms

Regulatory changes in 2017 and 2019 marked a turning point in Turkey's crowdfunding ecosystem. Licensing requirements, investor protection guidelines, and fund utilization standards legitimized equity-based crowdfunding, allowing platforms to thrive. Fonbulucu emerged as a market leader, managing 56% of funds raised in 2021–2023, with over 189 projects collectively raising approximately 1 trillion Turkish Lira (1).

### 5.2. Integration of Blockchain-Based Systems

Blockchain integration addresses longstanding inefficiencies, including transparency, costs, and liquidity. Immutable ledgers ensure contributors can monitor fund utilization in real time, building stakeholder trust. Smart contracts automate processes such as KYC / AML compliance, milestone-based payouts, and refunds, significantly reducing operational costs. Tokenization allows fractional ownership and secondary market trading, enabling investors to exit positions early and enhancing platform liquidity.

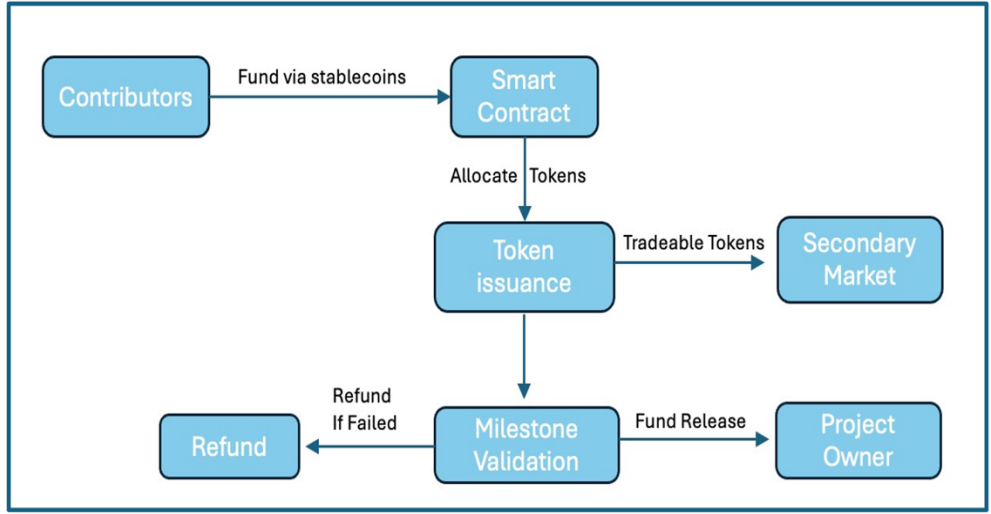

Figure 3: Blockchain Workflow for Equity Crowdfunding highlighting smart contract and tokenization processes (created by author).

### 5.3. Sector-Specific Insights

Turkish equity crowdfunding platforms predominantly serve the information technology, health technologies, and gaming sectors. These industries leverage crowdfunding for scalable and innovative projects, aligning with the capabilities of Blockchain-based systems. For instance, health technology startups use crowdfunding to develop medical devices, while gaming projects attract contributions through tokenized equity shares.

### 5.4. Challenges and Solutions

Although promising, the Turkish market faces challenges specific to Blockchain adoption:

- **Regulatory Ambiguity:** Collaboration with financial regulators is required to establish transparent compliance standards for Blockchain and stablecoin usage.
- **User Education:** Limited Blockchain literacy among contributors and creators necessitates educational resources such as tutorials and webinars.
- **Stablecoin Access:** Partnerships with local exchanges like BtcTurk simplify stablecoin acquisition, ensuring ease of use for contributors.

### 5.5. Future Outlook

Integrating Blockchain technology into Turkey's crowdfunding platforms will enhance global competitiveness. Key developments include:

- The launch of secondary markets for token trading, enabling investor liquidity.
- Increased adoption of stablecoins for cross-border fundraising.
- Continued collaboration between industry stakeholders and regulators to foster innovation while ensuring compliance.

## 6. Conclusion

The proposed Blockchain-based crowdfunding framework represents a significant leap forward in addressing the inefficiencies of traditional platforms. This framework offers a transformative approach to cost reduction, transparency, scalability, and investor trust by integrating stablecoins, smart contracts, and tokenization into a decentralized model.

*Key Findings* This study identifies several critical advantages of the Blockchain-based framework. Traditional crowdfunding models burden users with high fees, lack of transparency, and restricted liquidity. In contrast, Blockchain technology enables peer-to-peer transactions that bypass intermediaries, reducing operational costs (10). Smart contracts enforce predefined conditions, such as milestone-based fund disbursement, ensuring funds are transparent and securely managed (2). Tokenization introduces liquidity to crowdfunding, enabling contributors to trade tokens on secondary markets, thereby reducing the financial risks associated with locked capital. Additionally, stablecoins facilitate seamless global transactions, eliminating currency conversion complexities and enhancing accessibility for contributors worldwide (6).

*Implications for Crowdfunding* Adopting this framework can reshape the crowdfunding landscape by addressing long-standing challenges. Entrepreneurs, especially in emerging markets, can benefit from reduced costs and democratized access to capital. Tokenization empowers micro-investments, opening avenues for smaller-scale projects previously excluded from traditional funding mechanisms. Furthermore, the integration of stablecoins extends the global reach of campaigns, fostering economic activity in regions with limited access to traditional funding sources (11).

*Challenges and Future Directions* Despite its potential, several challenges must be addressed for widespread adoption. Regulatory ambiguity surrounding Blockchain and cryptocurrencies may hinder implementation in certain jurisdictions. Technological barriers, such as user unfamiliarity with Blockchain systems, necessitate robust educational initiatives. Stablecoin accessibility in emerging markets and resistance from traditional platforms are additional hurdles. Future framework iterations could explore integration with Decentralized Finance (DeFi) platforms to offer contributors staking and yield opportunities. Piloting the framework in diverse regulatory environments and collaborating with local exchanges will enhance usability and scalability.

*Final Remarks* This framework offers a paradigm shift in equity crowdfunding, leveraging Blockchain technology to set new transparency, efficiency, and investor trust standards. While challenges remain, the benefits outweigh the hurdles, positioning this model as a promising solution for the future of entrepreneurial financing. With continued innovation and collaboration, Blockchain-based crowdfunding has the potential to drive sustainable economic growth and foster global investment.

# References

1. Mustafa Savaş Ünsal, *Equity Crowdfunding: Empowering Entrepreneurship Through Innovative Funding: Focus on the Turkish Market and a Blockchain Implementation Proposal*. Retrieved from Amazon.com, 2024.

2. Bhabendu Kumar Mohanta, Soumyashree S Panda, and Debasish Jena, *An overview of smart contracts and use cases in Blockchain Technology*, 9th International Conference on Computing, Communication, and Networking Technologies (ICCCNT), Bengaluru, India, 2018, pp. 1-4, doi: 10.1109/ICCCNT.2018.8494045, (2018).

3. Armin Schwienbacher and Benjamin Larralde, *Crowdfunding of small entrepreneurial ventures*, Social Science Research Network Electronic Journal, doi: 10.2139/ssrn.1699183, and in Handbook of Entrepreneurial Finance Oxford University Press, 2010.

4. Ethan Mollick, *The dynamics of crowdfunding: An exploratory study*, Journal of Business Venturing, Volume: 29, Issue: 1, pp. 1–16, doi: 10.1016/j.jbusvent.2013.06.005, 2014.

5. Huasheng Zhu and Zach Zhizhong Zhou, *Analysis and outlook of applications of Blockchain technology to equity crowdfunding in China*, Open access Financial Innovation, Volume: 2, Issue: 29, doi: 10.1186/s40854-016-0044-7, 2016.

6. Vitalik Buterin, *A Next-Generation Smart Contract and Decentralized Application Platform*, Ethereum Whitepaper, 2014.

7. Mahdi M. Miraz and Maaruf Ali, *Applications of Blockchain Technology beyond Cryptocurrency*, Article in Annals of Emerging Technologies in Computing, Volume: 2, No: 1, pp. 1-6, doi: 10.33166/AETiC.2018.01.001, 2018. arXiv:1801.03528v1[cs.CR].

8. Marc Pilkington, *Blockchain technology: Principles and applications*, Social Science Research Network Electronic Journal: Available, in Research Handbook on Digital Transformations edited by F. Xavier Olleros and Majlinda Zhegu. Edward Elgar, 2016.

9. Antunes, R., Guarda, T., *A systematic review of Blockchain use cases in crowdfunding*, Journal of Applied Business Research, 36(4), 165–177, 2020.

10. Javier Ramos, *Crowdfunding and the role of managers in ensuring the sustainability of crowdfunding platforms*, Report European 26596 EN doi: 10.13140/RG.2.1.3925.6480, 2014.

11. Satoshi Nakamoto, *Bitcoin: A Peer-to-Peer Electronic Cash System*, Retrieved from https://bitcoin.org/bitcoin.pdf, 2008.

12. Mustafa Savaş Ünsal, *Yeni Nesil İş Yapma Modeli Olarak Kitlesel Fonlama Ve Destekçi Motivasyonu*, Philosophiae Doctor dissertation, Istanbul Nişantaşı University, 111-112, 2023.